\documentclass[prb,twocolumn,showpacs]{revtex4-1}
\usepackage{epsfig}
\usepackage{graphicx}
\usepackage{color}

\begin{document}

\newcommand{\im}{{\rm Im}}
\newcommand{\re}{{\rm Re}}
\newcommand{\up}{\uparrow}
\newcommand{\dn}{\downarrow}
\newcommand{\eps}{\varepsilon}
\newcommand{\ave}[1]{\langle #1\rangle}
\newcommand{\iv}{\mathbf i}

\newcommand{\jpsj}{J.\ Phys.\ Soc.\ Jpn.}

\title{Finite-size effects in transport data from Quantum Monte Carlo simulations}

\author{Rubem Mondaini,$^1$ K. Bouadim,$^2$ Thereza Paiva,$^1$ and Raimundo R. dos Santos$^1$}
\affiliation{$^1$Instituto de Fisica, Universidade Federal do Rio de Janeiro Cx.P. 68.528, 21941-972 Rio de Janeiro RJ, Brazil\\ $^2$Department of Physics, Ohio State University; 191 West Woodruff Ave
Columbus OH 43210-1117, USA}

\begin{abstract}
{We have examined the behavior of the compressibility, the dc-conductivity, the single-particle gap, and the Drude weight as probes of the density-driven metal-insulator transition in the Hubbard model on a square lattice.
These quantities have been obtained through determinantal quantum Monte Carlo simulations at finite temperatures on lattices up to $16\times 16$ sites.
While the compressibility, the dc-conductivity, and the gap are known to suffer from `closed-shell' effects due to the presence of artificial gaps in the spectrum (caused by the finiteness of the lattices), we have established that the former tracks the average sign of the fermionic determinant ($\ave{sign}$), and that a shortcut often used to calculate the conductivity may neglect important corrections.
Our systematic analyses also show that, by contrast, the Drude weight is not too sensitive to finite-size effects, being much more reliable as a probe to the insulating state.
We have also investigated the influence of the discrete imaginary-time interval ($\Delta\tau$) on $\ave{sign}$, on the average density ($\rho$), and on the double occupancy ($d$): we have found that $\ave{sign}$ and $\rho$ are more strongly dependent on $\Delta\tau$ away from closed-shell configurations, but $d$ follows the $\Delta\tau^2$ dependence in both closed- and open-shell cases.
}
\end{abstract}
\pacs{
71.10.Fd    
71.30.+h	 
71.27.+a    
73.63.-b	 
}
\maketitle

\section{Introduction}
\label{sec:intro}

Metal-insulator transitions (MIT) are still a topic of intense activity.\cite{Imada98}
In clean systems, an otherwise metallic system can become an insulator through the opening of a gap in the spectrum due to electronic repulsion; they become what are known as Mott insulators.\cite{Mott78}
Alternatively, band insulators correspond to systems in which the valence band is completely filled, even in the absence of repulsive interactions.
When the on-site energies are different (but regularly distributed), due to, say different atomic species, electrons may become trapped: in this case the system is a charge-transfer insulator.
In addition, in the presence of disorder the system may become an insulator as a result of electrons being unable to diffuse throughout the lattice; i.e., they may undergo an Anderson localization transition.
One clear experimental signature of the insulating state is a vanishing conductivity as the temperature is decreased.
However, from the theoretical point of view, and in the context of Quantum Monte Carlo (QMC) simulations\cite{Blankenbecler81,Hirsch85,White89b,Loh90,dosSantos03} in particular, detecting an insulating state is not always straightforward.
First, one necessarily deals with systems of finite size, hence with gaps in the spectrum which may be of the same magnitude as the ones responsible for the insulating behavior.
These gaps occur at filling factors corresponding to `closed shells', and give rise to atypical behavior in several quantities of interest; further, these closed-shell effects, which are readily seen in the non-interacting case, can persist in the presence of interactions (see below).
Secondly, QMC simulations are plagued by the `minus-sign problem',\cite{Loh90,dosSantos03} which precludes the study of several low-temperature properties of the system as the electronic density is varied continuously.
And, finally, in spite of the wide variety of quantities at our disposal to probe a MIT, such as the compressibility, the dc-conductivity, the single-particle gap, and the Drude weight,\cite{Scalapino93} to name a few, they yield conflicting information in some cases, the origin of which is still not fully understood.
For instance, under somewhat restrictive conditions\cite{Randeria92,Trivedi95,Trivedi96,Scalettar99,Chakraborty07} the dc-conductivity can be calculated in a convenient way, without resorting to analytic continuation of imaginary-time QMC data to real frequencies, which may be a delicate matter;\cite{Jarrell96,Sandvik98} however, in the case of the Hubbard model, for some particular combinations of lattice size and electronic densities (away from half filling), the conductivity behaves as if the system were insulating, which casts doubts on whether the conditions are really met, or if it is a manifestation of `closed-shell' effects, or both.
In the case of homogeneous versions of well studied models, one may be able to generate data for many different lattice sizes for a given electronic density (`minus-sign problem' permitting); in this way, a trend with system size can be established, and any deviation from it should be readily identified.
However, this may not be the case of systems with an overlying structure, such as a superlattice,\cite{Paiva96,Paiva98} a checkerboard lattice, or even in the presence of staggered on-site energies (the ionic Hubbard model).\cite{Hubbard81,Paris07,Bouadim07,Merino09}

Our purpose here is to shed light into these discrepancies, and to compare different approaches to detect a MIT from QMC data; as a by-product, we will also establish a connection between the behavior of the compressibility and the infamous sign problem of the fermionic determinant.
The layout of the paper is as follows.
In Sec.\ \ref{sec:m&m} we introduce the Hubbard model, and outline the computational approach used.
In Sec.\ \ref{sec:kappa} we discuss the predictions from the electronic compressibility, when the effects of closed shells manifest themselves as a major finite-size effect.
Section \ref{sec:cond_dos} is devoted to finite-size effects on the dc-conductivity and the density of states, as obtained through an inverse Laplace transform of the current-current correlation function and the the single particle Green's function respectively; in this Section we also provide numerical estimates for the errors involved when the dc-conductivity is calculated setting the imaginary-time $\tau=\beta/2$, where, as usual, $\beta\equiv 1/T$, in units such that the Boltzmann constant is unity.
In Sec.\ \ref{sec:drude}, we discuss the Drude weight in detail, and show that it does not suffer from closed-shell effects.
The single-particle excitation gap is considered in Sec.\ \ref{sec:gap}, and we find that it suffers from the same closed-shell effects as the other probes of the insulating state, apart from the Drude weight.
In Sec.\ \ref{sec:sign} a systematic study leads to a connection between the sign of the fermionic determinant and the compressibility; we also discuss the influence of the imaginary-time interval on some of the data.
And, finally, Sec.\ \ref{sec:concl} summarizes our findings.

\section{Model and Calculational Details}
\label{sec:m&m}

The simplest model to capture the physics of Mott insulators is the repulsive Hubbard model, which is characterized by the Hamiltonian
\begin{eqnarray}
\mathcal{H} &=& - t \sum_{\langle i,j\rangle,\sigma} (c^\dagger_{i\sigma} c_{j \sigma}^{\phantom\dagger} + c^\dagger_{j\sigma} c_{i \sigma}^{\phantom\dagger} ) \nonumber
\\
&&+ U\sum_i \left(n_{i\uparrow}-\frac{1}{2}\right)\left(n_{i\downarrow}-\frac{1}{2}\right) -\mu \sum_i n_i, \label{eq:hamil}
\end{eqnarray}
where, in standard notation, $c_{i \sigma}$ is the fermion destruction operator at site $i$
with spin $\sigma=\uparrow,\downarrow$, $n_{i\sigma}=c^\dagger_{i\sigma}c^{\phantom{\dagger}}_{i\sigma}$, and $n_i=n_{i\uparrow}+n_{i\downarrow}$.
We only consider nearest-neighbor hopping (indicated by $\langle i,j\rangle$) on a two-dimensional $L\times L$ square lattice, and work in the grand-canonical ensemble; the chemical potential
$\mu$ is tuned to yield the desired density $\rho = \sum_i\langle n_i \rangle/N$, where $N=L^2$ is the number of lattice sites.
The hopping parameter $t$ sets the energy scale, so we take $t= 1$; throughout this paper, we have considered the weak- to intermediate coupling regime, $U\leq4$, for which size effects are more severe.

\begin{figure}[t]
\includegraphics[width=3.5in,angle=0]{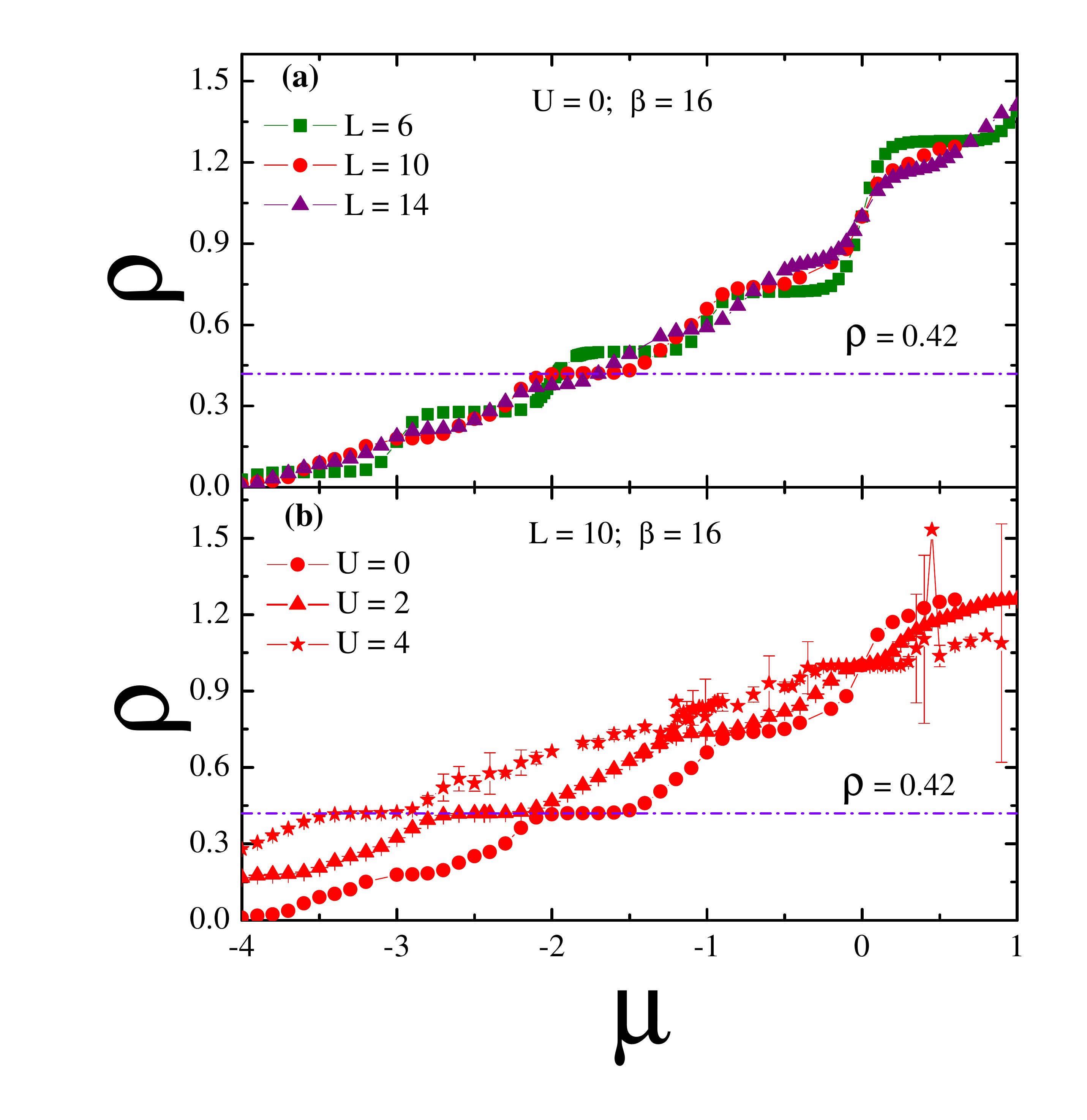}
\vspace{-0.2cm}
\caption{
(Color online) Electronic density vs.\ chemical potential: in (a) for the free case, at $\beta=16$, and for different linear lattice sizes $L$; in (b) for the $L=10$ lattice and different interactions $U$. The horizontal dashed lines highlight the specific density ($\rho=0.42$) at which one plateau appears for the $L=10$ lattice.
\vspace{-0.5cm}
}
\label{fig:electr-compress-u0}
\end{figure} 

We use determinant quantum Monte Carlo (DQMC) simulations\cite{Blankenbecler81,Hirsch83,Hirsch85,White89b,dosSantos03} to investigate the properties of the Hubbard model.
In this method, the partition function is expressed as a path integral by using the Suzuki-Trotter decomposition of $\exp(-\beta\mathcal{H})$, introducing the imaginary-time interval $\Delta\tau$.
The interaction term is decoupled through a discrete Hubbard-Stratonovich transformation,\cite{Hirsch83} which introduces an auxiliary Ising field.
This allows one to eliminate the fermionic degrees of freedom, and the summation over the auxiliary field (which depends on both the site and the imaginary time) is carried out stochastically.
Initially this field is generated randomly, and a local flip is attempted, with the acceptance rate given by the Metropolis algorithm.
The process of traversing the entire space-time lattice trying to change the auxiliary field variable constitutes one DQMC sweep.
 For most of the data presented here, we have used typically 1,000 warmup sweeps for equilibration, followed by 4,000 measuring sweeps, when the error bars are estimated by the statistical fluctuations; when necessary, the data were estimated over an average of simulations with different random seeds.
Typically, we have set $\Delta\tau=0.125$, but often data were also collected for $\Delta\tau=0.0625$, just to confirm that systematic errors are indeed small;  further, for some quantities we have also performed extrapolations towards $\Delta\tau\to 0$ from up to eight distinct values of $\Delta\tau$.
One should also keep in mind that since we do not use a checkerboard breakup of the lattice, our equal imaginary-time data for $U=0$ are exact, so that they do not depend on the imaginary-time discretization; the $\tau$-dependent quantities result from sampling even for $U=0$, but the statistical errors are negligible in this case.
With the updating being carried out on the Green's functions,\cite{Blankenbecler81,Hirsch85,dosSantos03} at the end of each sweep we have at our disposal both equal-`time' and $\tau$-dependent quantities, which we discuss in turn.

\begin{figure}[t]
\includegraphics[width=3.5in,angle=0]{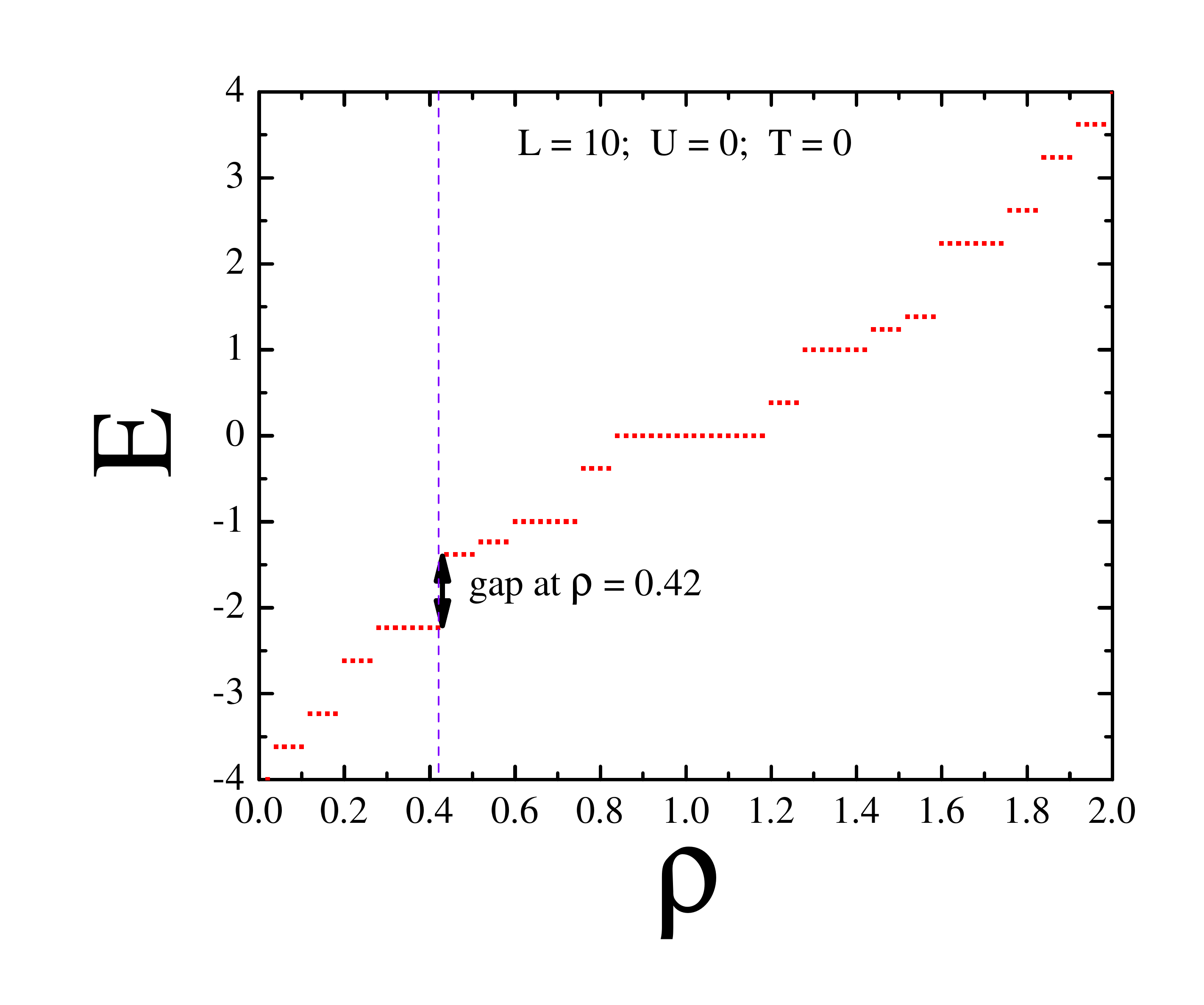}
\vspace{-0.2cm}
\caption{
(Color online) Total energy $E$ as a function of electronic density $\rho$ for the $L = 10$ lattice at $T=0$ in the noninteracting limit ($U=0$).
\vspace{-0.5cm}
}
\label{fig:Evsrho} 
\end{figure}

\section{Electronic Compressibility}
\label{sec:kappa}

Let us first consider the electronic compressibility, $\kappa=\rho^{-2}\partial\rho/\partial\mu$.
Being a direct measure of the charge gap, it may be used to detect insulating phases; a major computational advantage is that it is a local quantity, thus fluctuating very little within the DQMC approach.
In Fig.~\ref{fig:electr-compress-u0}(a), the density $\rho$ is plotted as a function of the chemical potential, for different lattice sizes, for the free case, $U=0$, and at a fixed temperature.
If taken at face value, plateaus in the $\rho\times\mu$ curves would be identified with incompressible phases, and hence with insulating regions.
However, a closer look reveals that both the width of the plateaus, as well as their positions, are strongly dependent on the finite system sizes used.
Given that for $U=0$ the system is certainly metallic for all densities, the presence of these plateaus can be traced back to gaps in the energy spectrum of the noninteracting Hubbard model on a finite square lattice, which is given in the usual way by $E=\sum_{{\bf q}\leq {\bf q}_F(\rho);\sigma} \varepsilon({\bf{q}})$, with $\varepsilon({\bf{q}})=-2t(\cos{q_x}+\cos{q_y})$, where ${\bf q}_F(\rho)$ is the Fermi wave vector for the density $\rho$.
In Fig.~\ref{fig:Evsrho} the total energy is shown as a function of the electronic density for a $10\times10$ lattice: the energy gaps do not have the same magnitude, and one should notice, in particular, the gap at $\rho=0.42$, which is quite large in comparison with the ones between levels with $E< -2$.
This gap appears as a plateau in the data for the $10\times 10$ lattice in Fig.~\ref{fig:electr-compress-u0}(a), indicated by the horizontal dashed line.
The existence of this `gap' is a manifestation of what is referred to as `the closed-shell problem' and is characteristic of the finiteness of the lattice.
It should be stressed that such effects are still present when the interaction is switched on, at least up to intermediate values of $U$: from Fig.\ \ref{fig:electr-compress-u0}(b), we see that the gap moves towards smaller values of $\mu$ as $U$ is increased, though without any noticeable decrease in magnitude; in what follows we illustrate further consequences of these closed-shell effects.
As the lattice size is increased, the gaps become smaller, and the plateaus in the electronic density become narrower, until they completely vanish in the bulk limit, $L\rightarrow\infty$. For this reason, from now on we will refer to these plateaus as pseudo-insulating states.
The use of the compressibility to locate insulating regions must therefore be supplemented with thorough analyses of the robustness and the width of the plateaus with system size and temperature.

\begin{figure}[t]
\vspace{-0.2cm}
\includegraphics[width=9.1cm,angle=0]{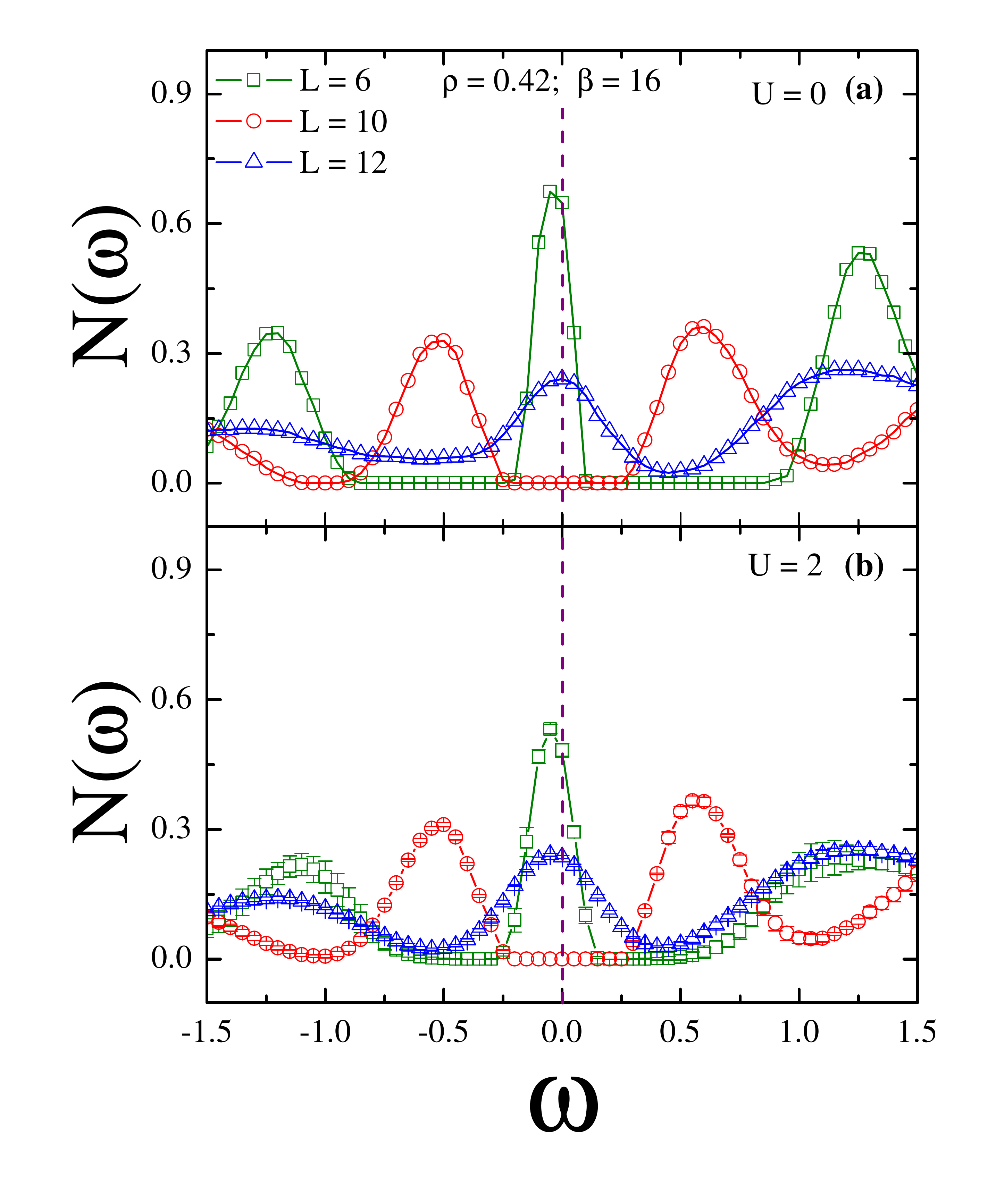}
\vspace{-1.0cm}
\caption{
(Color online) DOS spectrum for different lattice sizes at $\rho=0.42$, and for $U=0$ (a) and $U=2$ (b).
The Fermi energy ($\omega=0$) is shown as a dashed line.
The error bars represent statistical errors from different realizations.}
\label{fig:N(omega)vsomega} 
\end{figure}

\section{Conductivity and Density of States}
\label{sec:cond_dos}

The optical conductivity and the density of states (DOS) are other probes of the insulating state which are worth discussing in depth; this is especially in order, given that the use of the shortcut to calculate the dc-conductivity (see below) has been increasingly widespread,\cite{Chakraborty11a,Chakraborty11b} even beyond QMC.\cite{Sentef11}

First we recall that the simulations yield imaginary-time quantities, such as the real-space single-particle Green's function,
\begin{equation}
G({\bf r}\equiv{\bf i}-{\bf j},\tau)=\langle c_{\textbf{i}\sigma}(\tau)c^{\dagger}_{\textbf{j}\sigma}(0)\rangle,
\ \ 0\leq\tau\leq\beta,
\label{eq:gf}
\end{equation}
and the current-current correlation functions,
\begin{equation}
\Lambda (\textbf{q},\tau) \equiv \langle j_x(\textbf{q},\tau)j_x(-\textbf{q},0)\rangle,
\label{Ldef}
\end{equation}
where $j_x(\textbf{q},\tau)$ is the Fourier transform of the `time'-dependent current-density operator, $j_x({\bf i},\tau)\equiv e^{{\cal H}\tau}j_x({\bf i})\,e^{-{\cal H}\tau}$, with
\begin{equation}
j_x({\bf i})=it \sum_{\sigma} \left(
       c_{{\bf i}+{\bf \hat{x}},\sigma}^\dagger c_{{\bf i},\sigma}^{\phantom\dagger}
       -
       c_{{\bf i},\sigma}^\dagger c_{{\bf i}+{\bf \hat{x}},\sigma}^{\phantom\dagger}
\right).
\label{jdef}
\end{equation}
Now, the fluctuation-dissipation theorem yields\cite{Doniach98}
\begin{equation}
\Lambda(\mathbf{q}=0,\tau)=\int_{-\infty}^{\infty}\frac{d\omega}{\pi}\frac{e^{-\omega\tau}}{1-e^{-\beta\omega}}\ {\rm Im}\,\Lambda(\mathbf{q}=0,\omega), \label{eq:lambda-qtau}
\end{equation}
and linear response theory implies\cite{Mahan00}
\begin{equation}
{\rm Im}\,\Lambda(\mathbf{q}=0,\omega)=\omega\,{\rm Re}\,\sigma(\omega);
\end{equation}
similarly, we have\cite{Doniach98,Mahan00}
\begin{equation}
G(\mathbf{r}=0,\tau)=\int_{-\infty}^{\infty} d\omega\ \frac{e^{-\omega\tau}}{1+e^{-\beta\omega}}\ N(\omega).
\label{eq:G-rtau}
\end{equation}
The calculation of $\sigma(\omega)$ and $N(\omega)$ is then reduced to numerically invert these Laplace transforms at a given temperature.
Here we employ an analytical continuation method,\cite{Sandvik98} through which the conductivity and the DOS can be obtained for the whole spectrum $\omega$. \cite {alpha}
While there has been some debate over which type of analytic continuation method is best suited to perform these Laplace transforms,\cite{Jarrell96,Syljuasen08} our purpose here is not to perform a systematic study of the outstanding issues; instead, we adopt one of the procedures\cite{Sandvik98} to extract estimates for $\sigma(\omega)$ which, in turn, will be used to test the trends in the calculation of $\sigma_{dc}$, as discussed below.

\begin{figure}[t]
\includegraphics[width=9.5cm,angle=0]{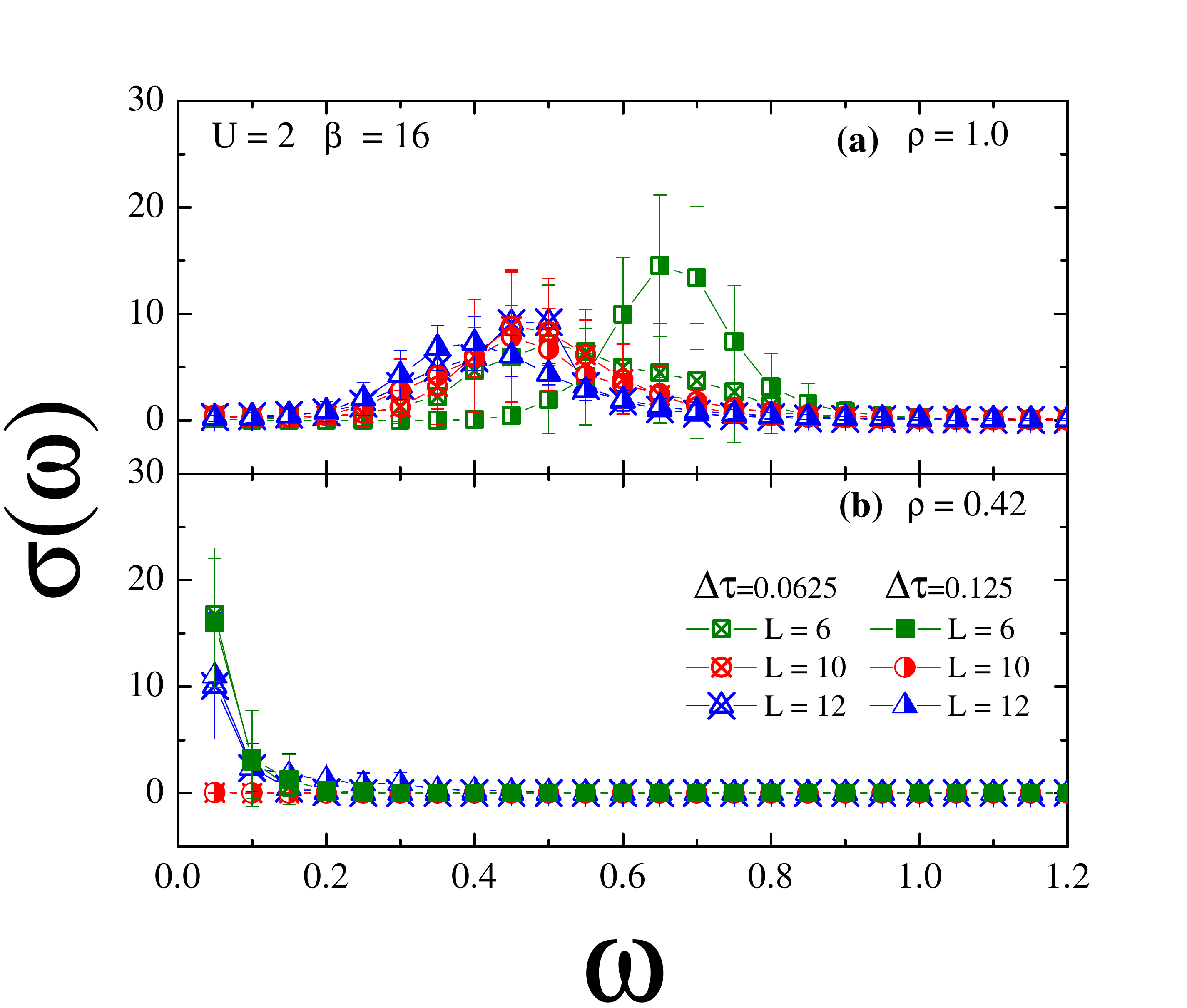}
\vspace{-0.5cm}
\caption{ (Color online) Optical conductivity from inverted Laplace transform (see text) at (a) half filling, $\rho=1$, and (b) for $\rho=0.42$, at given $U$ and inverse temperature, for different lattice sizes and $\Delta\tau$.
The error bars represent statistical errors from different realizations.}
\label{fig:sigma_vs_omega}
\end{figure}  

In Fig.\ \ref{fig:N(omega)vsomega} we compare the DOS at density $\rho=0.42$ for the free and interacting cases, obtained through the method described in Ref.\ \onlinecite{Sandvik98}.
It is clear that irrespective of the value of $U$, the DOS vanishes at the Fermi energy for $L=10$, while being non-zero for $L=6$ and $12$.
Figure \ref{fig:sigma_vs_omega} shows the optical conductivity for the interacting case ($U=2$), calculated with the same inversion method,\cite{Sandvik98} both at half filling, and for $\rho=0.42$.
We see that, while at half filling the insulating behavior is apparent for all system sizes [$\sigma_{\rm dc}(T)=\lim_{\omega\to0}\sigma(\omega,T)\to0$], for $\rho=0.42$ one would be led to identify an insulating behavior if only data for a $10\times 10$ lattice were available.
One should also note that data for both $\Delta\tau=0.125$ and 0.0625 are the same, within error bars.
The origin of this `false insulating' behavior can therefore be traced back to the closed shell
effects discussed above, though here $\sigma(\omega)$ is particularly affected by the large gap required to add an electron to the `closed shell' of 42 electrons; an analogous problem occurs at the
closed shell density of $\rho=86/144$ for the $12\times 12$ lattice (not shown).

In addition to suffering from the closed shell problem, the inversion procedure adopted\cite{Sandvik98} can be very costly in computer time, due to the need of very small error bars in the data for $\Lambda$.
An alternative method\cite{Randeria92,Trivedi95,Trivedi96,Scalettar99} to obtain $\sigma_{\rm dc}(T)$ consists of setting $\tau=\beta/2\equiv1/2T$ in Eq.\ (\ref{eq:lambda-qtau}), and assuming $\sigma(\omega)$ admits a Taylor
expansion near $\omega=0$; the integral can, in principle, be carried out term by term in the surviving even powers of $\omega$, and we get
\begin{equation}
\sigma_{\rm dc} (T)\approx \sigma_{\rm dc}^{(0)} (T)+ \sigma_{\rm dc}^{(2)}(T), \label{eq:sigma-dc1}
\end{equation}
plus higher order terms, with
\begin{equation}
\sigma_{\rm dc}^{(0)} (T)= \frac{1}{\pi T^2}\,\Lambda\left(\mathbf{q}=0,\tau=1/2T\right) \label{eq:sigma-dc0}
\end{equation}
and
\begin{equation}
\sigma_{\rm dc}^{(2)}(T)=-T^2\pi^2\left(\frac{\partial^2\sigma}{\partial\omega^2}\right)_{\omega=0}. \label{eq:sigma-dc2}
\end{equation}

\begin{figure}[t]
\vspace{-1.0cm}
\includegraphics[width=9.5cm,angle=0]{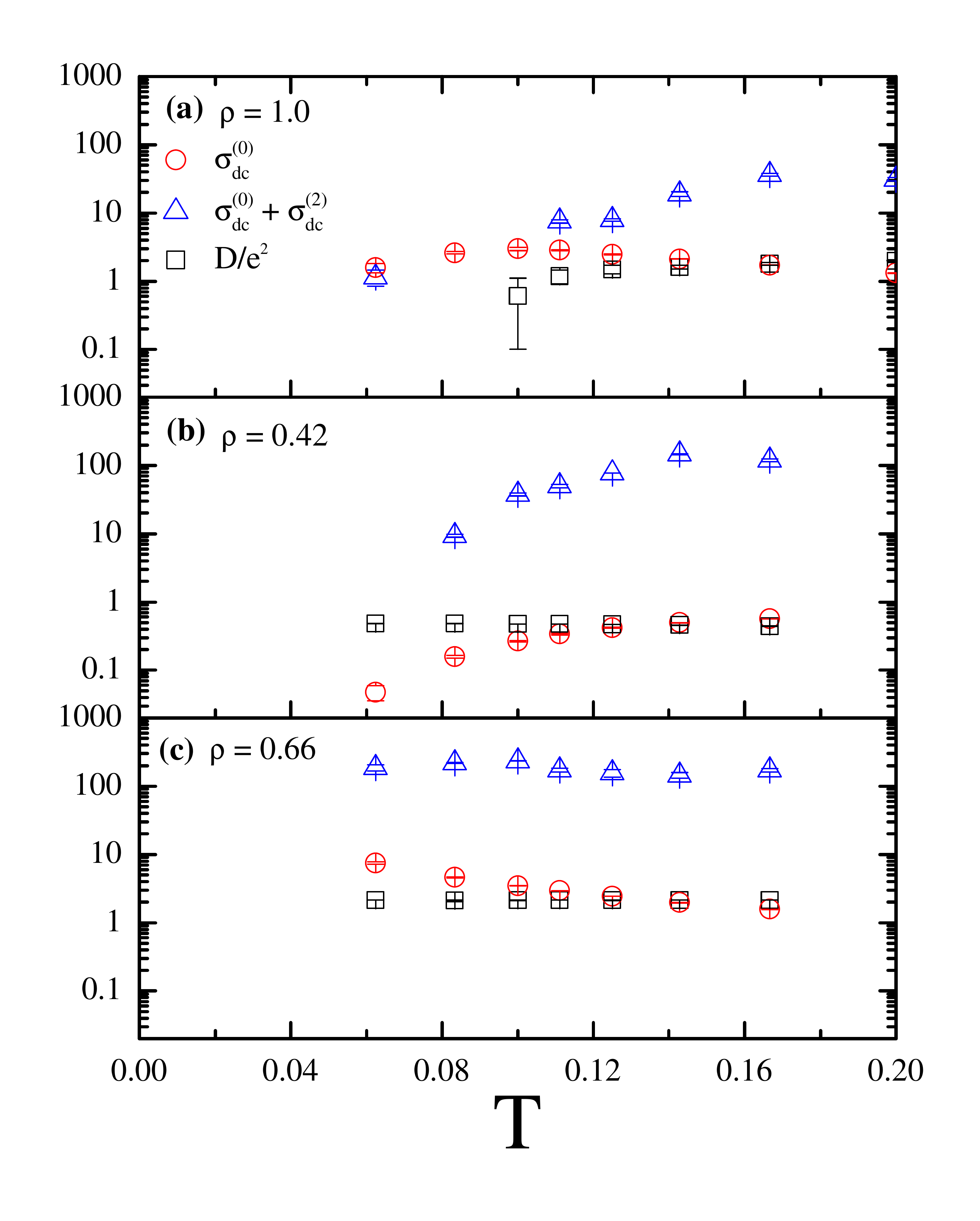}
\vspace{-0.5cm}
\caption{ (Color online) Temperature dependence of the dc-conductivity [circles, zeroth order in $\omega$, and triangles, up to second order; see Eqs.\ (\ref{eq:sigma-dc1})-(\ref{eq:sigma-dc2})], and of the Drude weight (squares; see Sec.\ \ref{sec:drude}), for $U=2$, on  a $10\times 10$ lattice, and for different electronic densities.
The error bars for $\sigma_{\rm dc}$ are due to the averaging process, while those for $D/e^2$ are due to extrapolations towards $\omega_m\to 0$.}
\label{fig:sigma-drude_T}
\end{figure}  

Note that if one wants to take $\sigma_{\rm dc}^{(0)}(T)$ as an approximation for $\sigma_{\rm dc}(T)$, $\sigma_{\rm dc}^{(2)}(T)$ must be small; this should occur if the temperature is low enough, and
the frequency dependence of the conductivity is smooth, i.e., if $T\ll \Omega$, where $\Omega$ sets a small energy scale of the problem.\cite{Trivedi95} While it is hard to assess {\it a priori} if
this condition is satisfied, in the present case we have data for $\sigma(\omega)$ at our disposal, for several temperatures; this allows us to calculate $\sigma_{\rm dc}^{(2)}(T)$, and check the
errors involved in neglecting it in Eq.\ (\ref{eq:sigma-dc1}). For $\rho=1$, we see from Fig.\ \ref{fig:sigma-drude_T}(a) that $\sigma_{\rm dc}^{(0)} (T)$ rises as the temperature is lowered [(red)
circles], but eventually bends down at some temperature, consistently with $\sigma_{\rm dc}^{(0)}\to0$ as $T\to 0$; in a generic situation, in which QMC data for these lowest temperatures were not
available, one could be misled to state that the system is metallic. However, when $\sigma_{\rm dc}^{(2)}(T)$ is included [(blue) triangles in Fig.\ \ref{fig:sigma-drude_T}(a)], the conductivity
acquires the correct steady decrease with decreasing $T\lesssim 0.15$; this shows that higher order terms may indeed be crucial at temperatures not so low. Figure \ref{fig:sigma-drude_T}(b) shows that
for $\rho=0.42$, $\sigma_{\rm dc}^{(0)}$ steady decreases as $T$ decreases, which is suggestive of insulating behavior; the inclusion of data for $\sigma_{\rm dc}^{(2)}(T)$ does not revert this trend.
Since for other densities the metallic behavior is unequivocal [see, e.g., Fig.\ \ref{fig:sigma-drude_T}(c)], one concludes that the spurious effect for $\rho=0.42$ is yet another manifestation of the
closed-shell density.
We have found that these overall features are also present for $U=4$; in particular, the contribution of $\sigma_{\rm dc}^{(2)}(T)$ when $\rho=0.42$, though significant, is again not sufficient to yield a metallic behavior, thus confirming that the false insulating state is indeed a closed shell effect.

From this analysis we conclude that extreme care must be taken when examining $\lim_{T\to0}\sigma_{\rm dc}^{(0)}(T)$ to indicate whether the ground state is metallic or
insulating; in addition, while the overall trend may be captured (away from closed-shell densities), attempts to fit experimental data with $\sigma_{\rm dc}^{(0)}(T)$ should lead to error, if the
temperatures involved are not too low.

\begin{figure}[t]
\includegraphics[width=3.5in,angle=0]{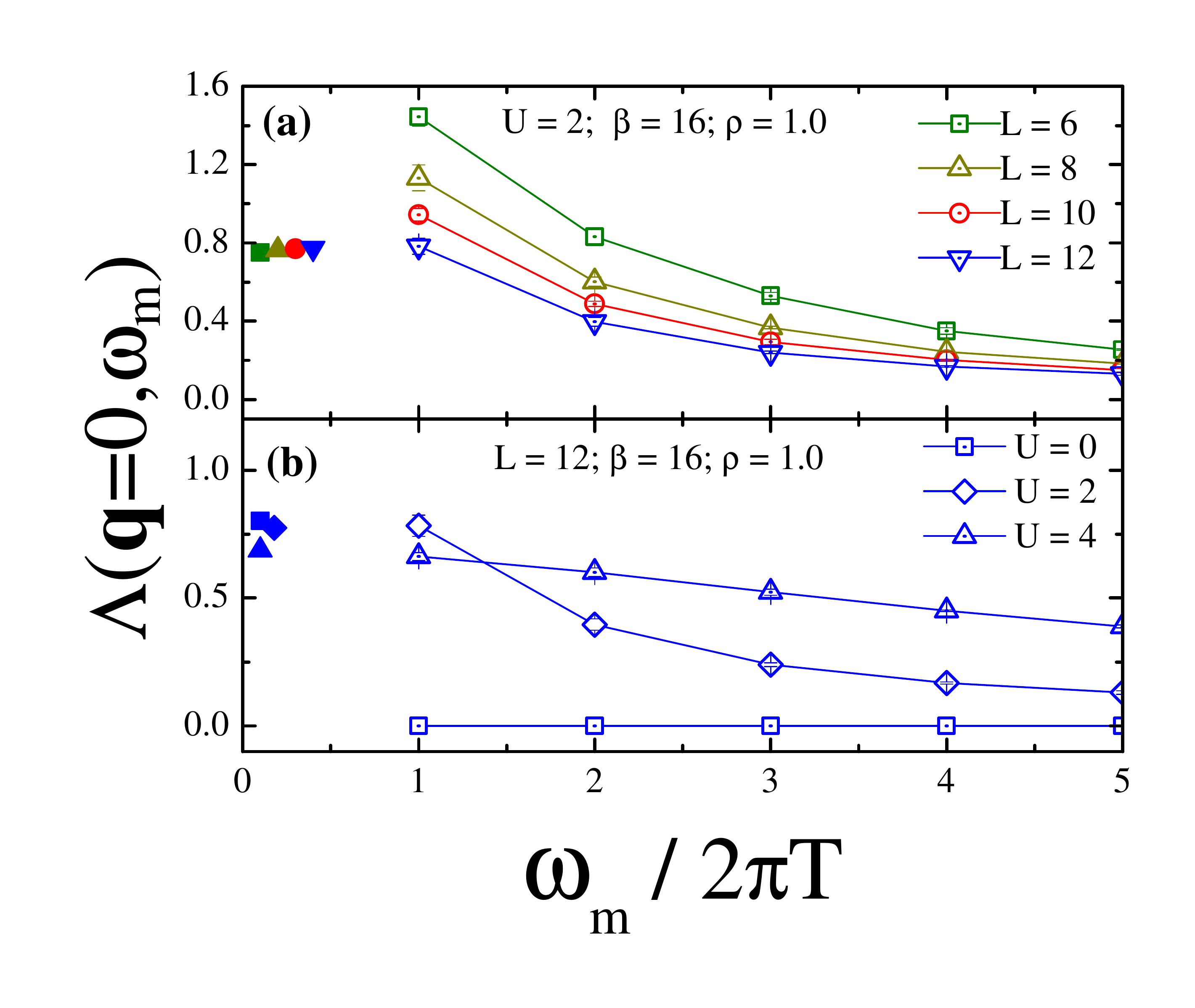}
\caption{(Color online)
Current-current correlation function $\Lambda(\textbf{q}=0,\omega_m)$ at half-filling $\rho=1.0~$, as a function of $\omega_m/2\pi T$, where $\omega_m$ is the Matsubara frequency, at a fixed inverse temperature $\beta=16$. The solid symbols denote $\langle-k_x\rangle$. In (a) the on-site repulsion is kept fixed and the data correspond to different linear lattice sizes; in (b) data are for a $12\times 12$ lattice, but for different values of $U$. The error bars represent statistical errors from different realizations.
}
\label{fig:Lambdarho1} 
\end{figure}

\section{The Drude Weight}
\label{sec:drude}

We now discuss the Drude weight, $D$, defined through
\begin{equation}
\lim_{T\to 0} \re\ \sigma(\omega,T) = D\, \delta(\omega) + \sigma_{\rm reg}(\omega), \label{eq:drude0}
\end{equation}
where $\sigma_{\rm reg}(\omega)$ is the regular (or incoherent) response.
Approximants to $D$ are readily available from QMC simulations as\cite{Scalapino93,Scalettar99}
\begin{equation}
\frac{{\tilde D}_m(T)}{\pi e^2}\equiv [\langle-k_x\rangle-\Lambda(\textbf{q}=0,i\omega_m)],
\label{eq:drude_approx}
\end{equation}
where $\omega_m=2m\pi T$ is the Matsubara frequency, and $\langle k_x\rangle$ is the average kinetic energy of the electrons per lattice dimension.
The Drude weight is then given by
\begin{equation}
\lim_{T,m\to 0} {\tilde D}_m \equiv \lim_{T\to 0} D(T)= D.
\label{eq:drude_lim}
\end{equation}
In actual calculations, both limits should be taken through extrapolations of sequences of low-temperature frequency-dependent data ${\tilde D}_m(T)$;\cite{Heidarian07} finite-size effects and finite-$\Delta\tau$ effects must also be taken into consideration when analyzing the data.

Figure \ref{fig:Lambdarho1}(a) illustrates how the uniform current-current correlation function at half filling depends on the Matsubara frequency, with both $\beta$ and $U$ fixed, for different system sizes. While $\ave{-k_x}$ is hardly dependent on the system size (see solid symbols in Fig.\ \ref{fig:Lambdarho1}), the same does not hold for $\Lambda({\mathbf 0},\omega_m)$.
Nonetheless, approximants to the Drude weight, as given by Eq.\ (\ref{eq:drude_approx}), do indeed approach zero with growing linear lattice size $L$, as it should for an insulating state.
Figure \ref{fig:Lambdarho1}(b) displays the same quantity, now for a fixed system size, but for different values of $U$; we see that as $m\to 0$, ${\tilde D}_m\to 0$ for $U\neq0$, while ${\tilde D}_m$ approaches a non-zero value for $U=0$.

\begin{figure}[t]
\includegraphics[width=3.5in,angle=0]{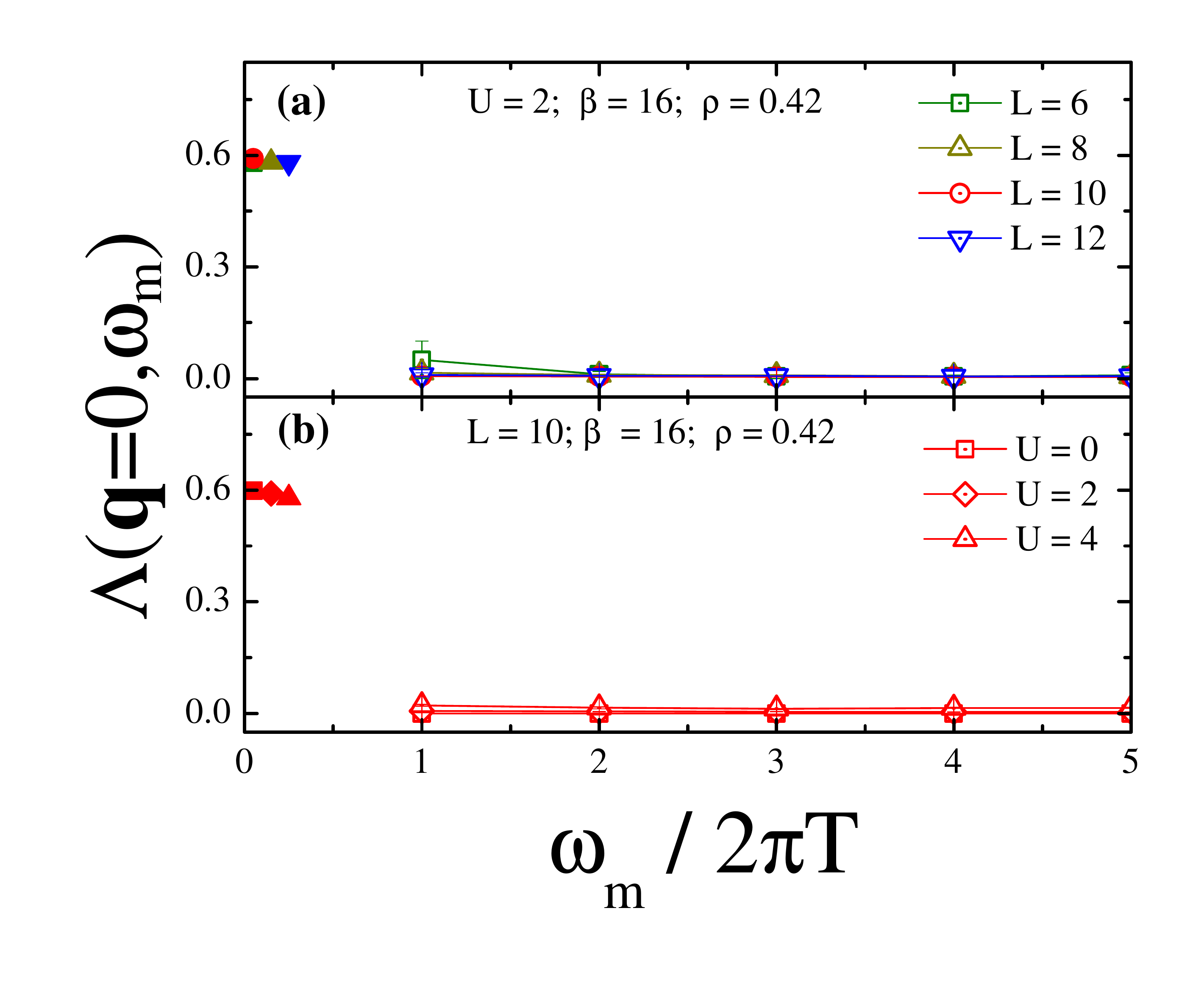}
\caption{(Color online)
Same as Fig.\ \ref{fig:Lambdarho1}, but for $\rho=0.42$; in (b) data are for a $10\times 10$ lattice.
}
\label{fig:Lambdarho042} 
\end{figure}

Data for $\rho=0.42$ and $U=2$ are shown in Fig.~\ref{fig:Lambdarho042}.
We see that ${\tilde D}_m/\pi e^2$ [Eq.\ (\ref{eq:drude_approx})] for the $10\times 10$ lattice does not show any false insulating behavior, as it did for other quantities: in (a) the difference between $\ave{-k_x}$ and $\lim_{\omega_m\to 0}\Lambda({\mathbf 0},\omega_m)$ does not display a significant change with lattice size, while in (b) the data show that the closed shell problem does not manifest itself over a wide range of values of $U$.

\begin{figure}[t]
\includegraphics[width=3.5in,angle=0]{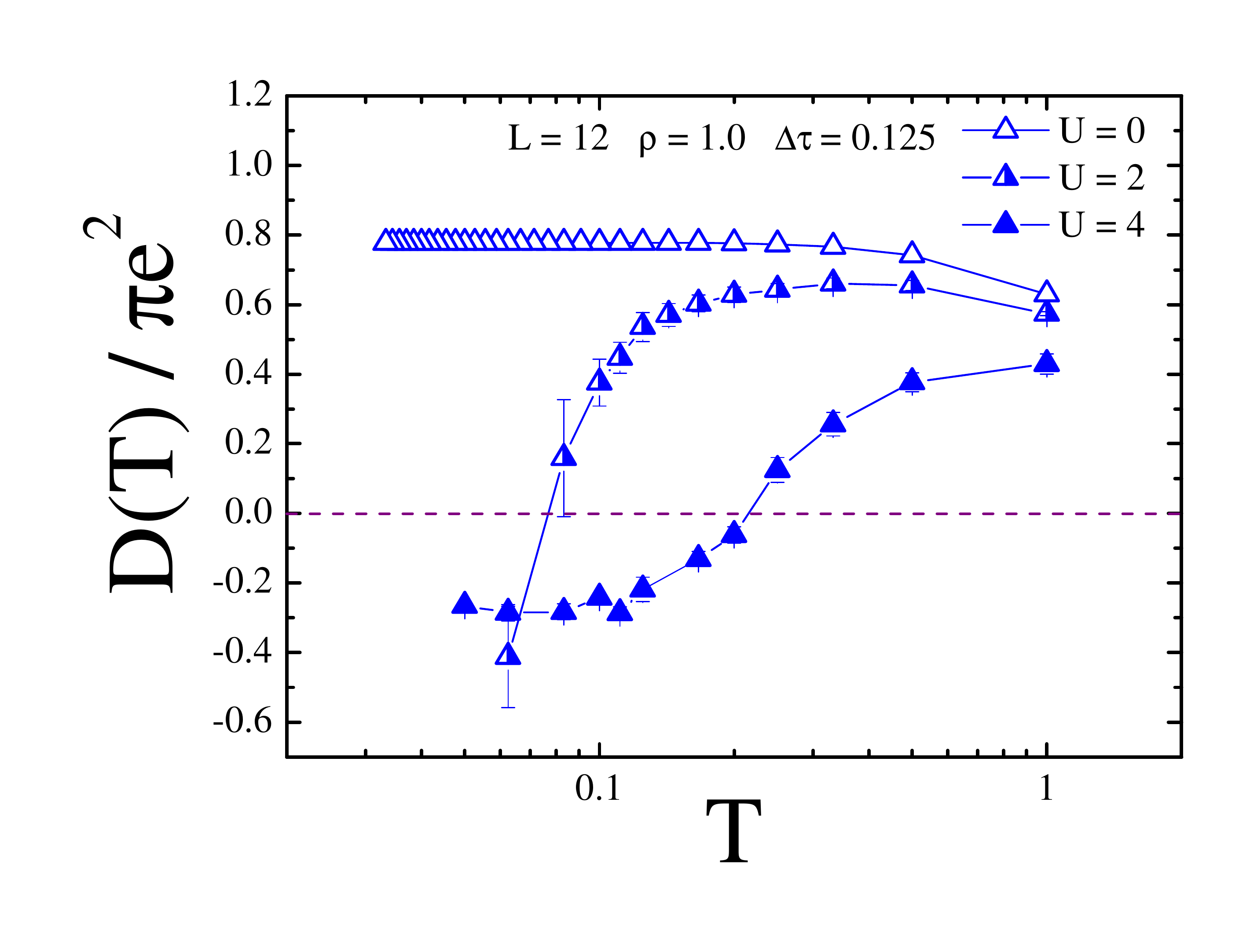}
\vspace{-0.2cm}
\caption{(Color online)
Drude weight approximants as a function of temperature for $\rho=1.0$ for a $L=12$ lattice for different values of $U$.
The error bars result from uncertainties in the extrapolations $\omega_m/2\pi T\to 0$; see text.
}
\label{fig:DvsTn1} 
\end{figure}

\begin{figure}[t]
\includegraphics[width=9.5cm,angle=0]{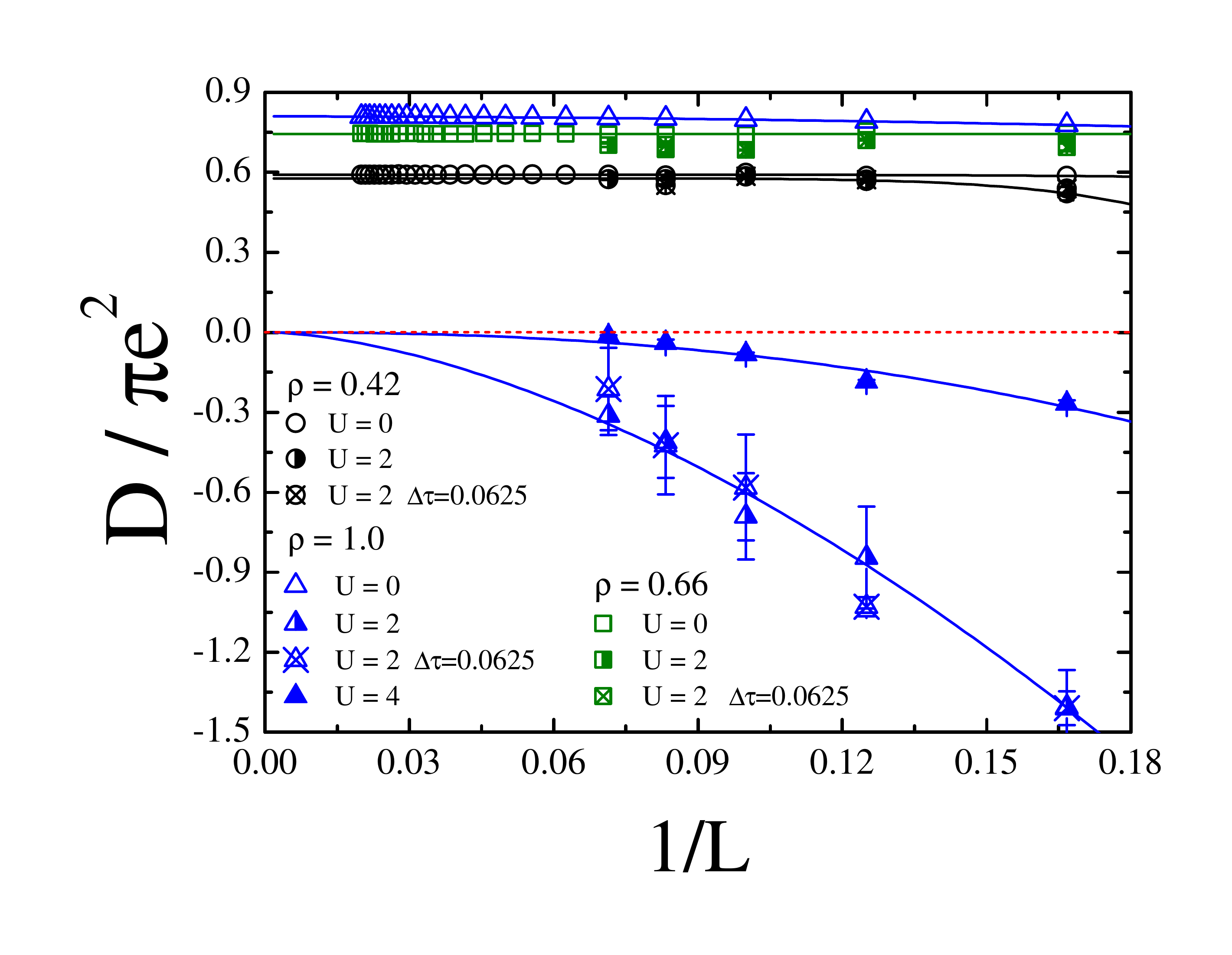}
\caption{(Color online) Size dependence of the (normalized) Drude weight at a fixed, finite temperature, $\beta=16$, for different densities: $\rho=0.42$ (circles), $\rho=0.66$ (squares), and $\rho=1$ (triangles). Empty,  half-filled, and filled symbols respectively correspond to $U=0$,  2, and 4; data are for $\Delta\tau=0.125$, except those with crossed symbols.
Error bars result from uncertainties in the extrapolations $\omega_m/2\pi T\to 0$ (see text) and are only appreciable for $\rho=1$.
 }
\label{fig:drude_L}
\end{figure}

In order to extract more quantitative data, we adopt the following procedure: For fixed $L$, $U$, and $\beta$, we plot ${\tilde D}_m$ as a function of $m\equiv\omega_m/2\pi T$, and extrapolate to $m\to 0$ with the aid of a parabolic fit to the data for the smallest $m$'s (figure not shown); we then obtain the temperature-dependent Drude weight, $D(T)$, appearing in Eq.\ (\ref{eq:drude_lim}).
By varying the temperature, system size, and $U$, we can generate plots of $D(T)$, examples of which are shown in Figs.\ \ref{fig:DvsTn1}  and \ref{fig:drude_L}.
As shown in Fig.\ \ref{fig:DvsTn1}, for a $12\times12$ lattice at half filling, in the non-interacting case the Drude weight clearly extrapolates to a non-zero value as $T\to 0$.
For $U>0$, $D(T)$ vanishes at some temperature $T_0(L,U)$, which increases with $U$ for a given $L$.
Data for half filling in Fig.\ \ref{fig:drude_L} show that at a fixed temperature, the Drude weight vanishes as the lattice size increases; that is, the points below $T_0$ in Fig.\ \ref{fig:DvsTn1} should approach the $D=0$ line for sufficiently large $L$.

%
\begin{figure}[t]
\vspace{-0.7cm}
\includegraphics[width=9.5cm,angle=0]{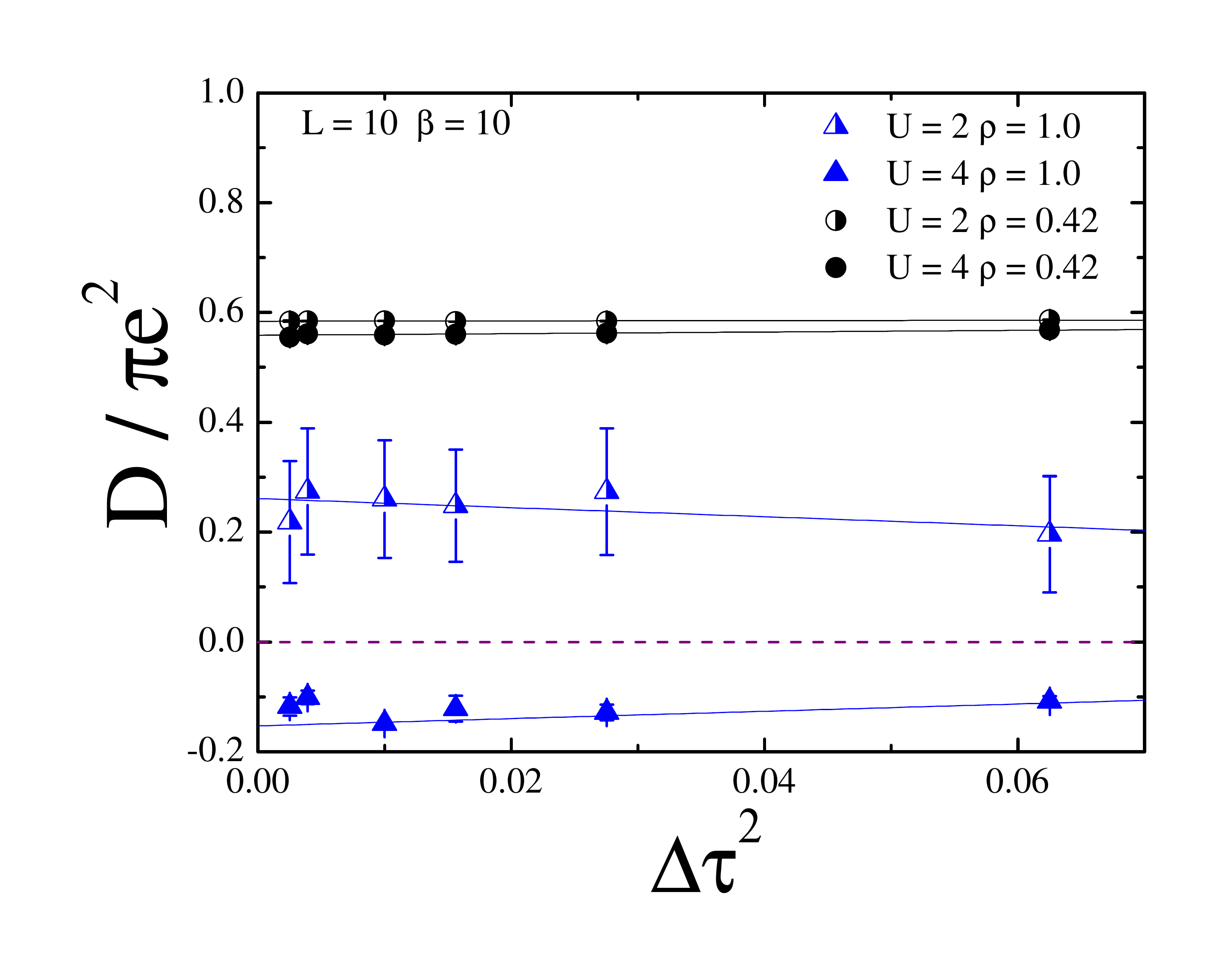}
\caption{ (Color online) Dependence of the (normalized) Drude weight with the  square of the `time' interval, at fixed temperature and lattice size at half filling (triangles), and at $\rho=0.42$ (circles).
Error bars result from uncertainties in the extrapolations $\omega_m/2\pi T\to 0$ (see text).
}
\label{fig:drude_dtau}
\end{figure}  

Away from half filling, the minus-sign problem prevents us from analyzing the size-dependence at very low temperatures, and we are restricted to data for $\beta=16$ for the densities $\rho=0.42$, and 0.66, while keeping $U\leq2$.
Nonetheless, some important conclusions can be drawn from our analyses of the data for $D(T)$ on finite-sized lattices: (1) we have found no evidence of a vanishing Drude weight at fixed, finite temperatures in the limit $L\to \infty$, as previously suggested for the one-dimensional case;\cite{Zotos01} (2) the dependence of $D$ with $1/L$, for fixed both temperature and on-site repulsion, is rather weak, without suffering from closed-shell effects, thus rendering extrapolations towards $L\to\infty$ trustworthy.
Once again, data for $\Delta\tau=0.125$ are the same as those for $\Delta\tau=0.0625$, within error bars.

Our results therefore show that the Drude weight has been hitherto unjustifiably overlooked as a reliable probe of the metal-insulator transition; its use should be more widespread, given that it is free from closed-shell effects, and its clear-cut `temperature-dependence' allows for an unambiguous characterization of insulating states.

\begin{figure}[t]
\includegraphics[width=3.5in,angle=0]{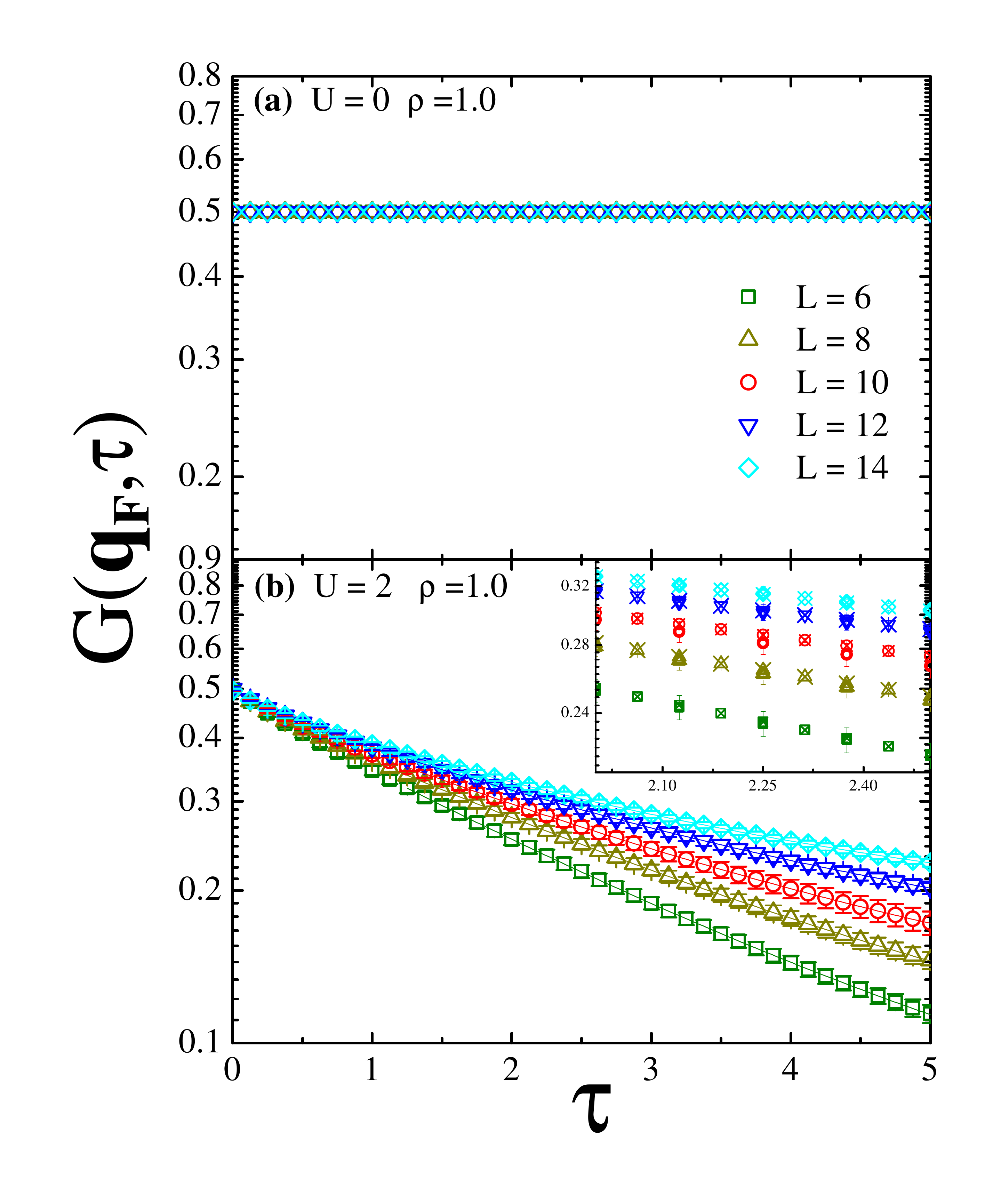}
\vspace{-0.2cm}
\caption{(Color online)
Log-linear plot of the imaginary-time dependence of the Green's function $G(\textbf{q},\tau)$ at the Fermi wave vector ${\mathbf q}_F$, for different lattice sizes at half filling, $\rho=1.0$, for the non-interacting (a), and interacting (b) cases.
The error bars in (b) are due to statistical errors from averaging over different realizations, and equivalent ${\bf q}_F$ points; here $\Delta\tau=0.125$.
The inset includes data obtained with $\Delta\tau=0.0625$, denoted by the corresponding crossed symbols from the main panel.
\vspace{-0.5cm}
}
\label{fig:G(q,tau)rho1} 
\end{figure}

\begin{figure}[t]
\includegraphics[width=3.5in,angle=0]{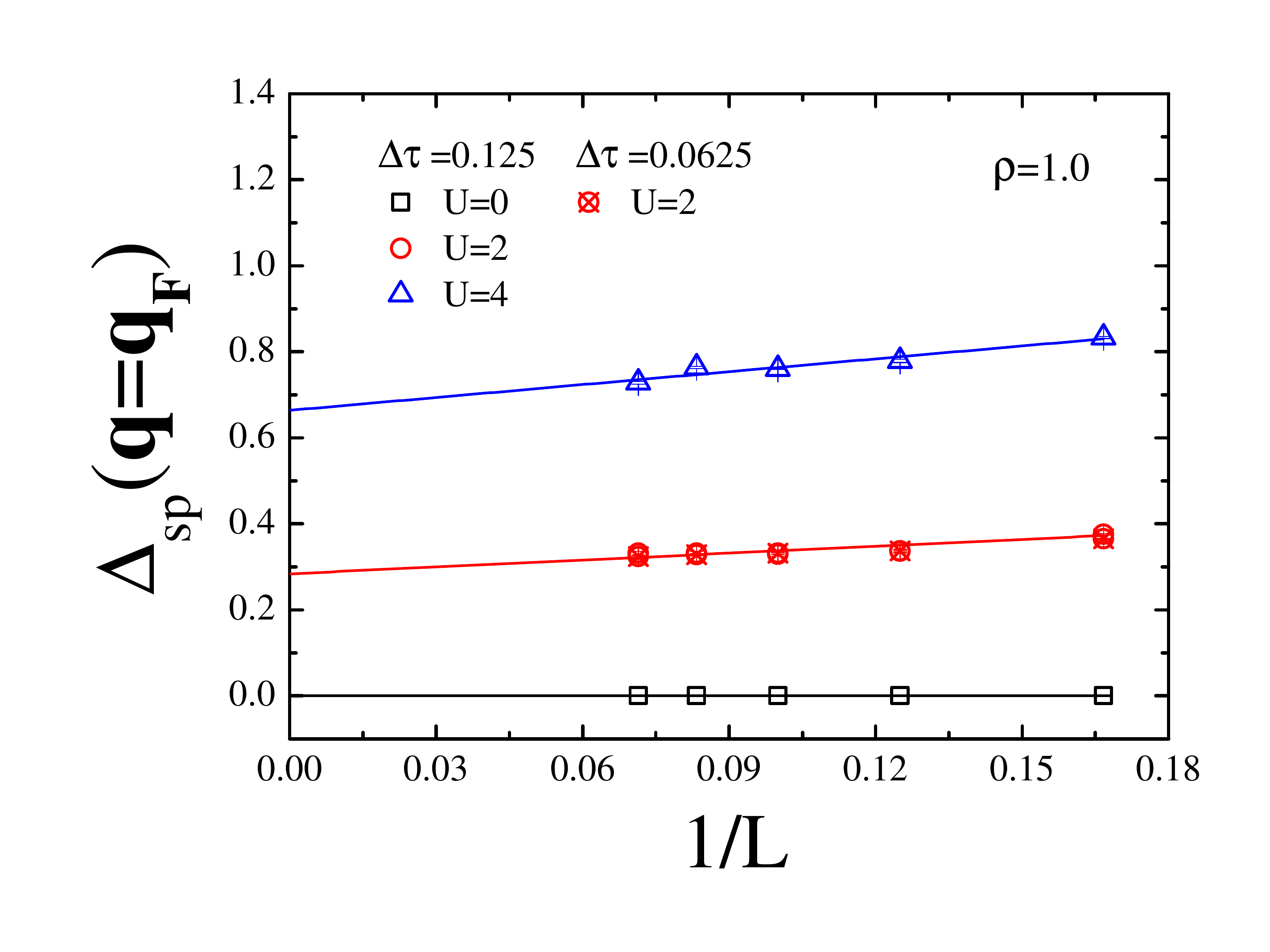}
\vspace{-0.8cm}
\caption{(Color online)
Finite-size dependence of the single particle excitation gap $\Delta_{sp}(\textbf{q}_F)$ at half-filling, for different values of the on-site repulsion.
The error bars are due to the exponential fits to the data for $G(\textbf{q}_F,\tau)$; see text.
The crossed symbols denote the corresponding data for $\Delta\tau$=0.0625.
}
\vspace{-0.3cm}
\label{fig:gapsprho1} 
\end{figure}
\begin{figure}
\includegraphics[width=3.5in,angle=0]{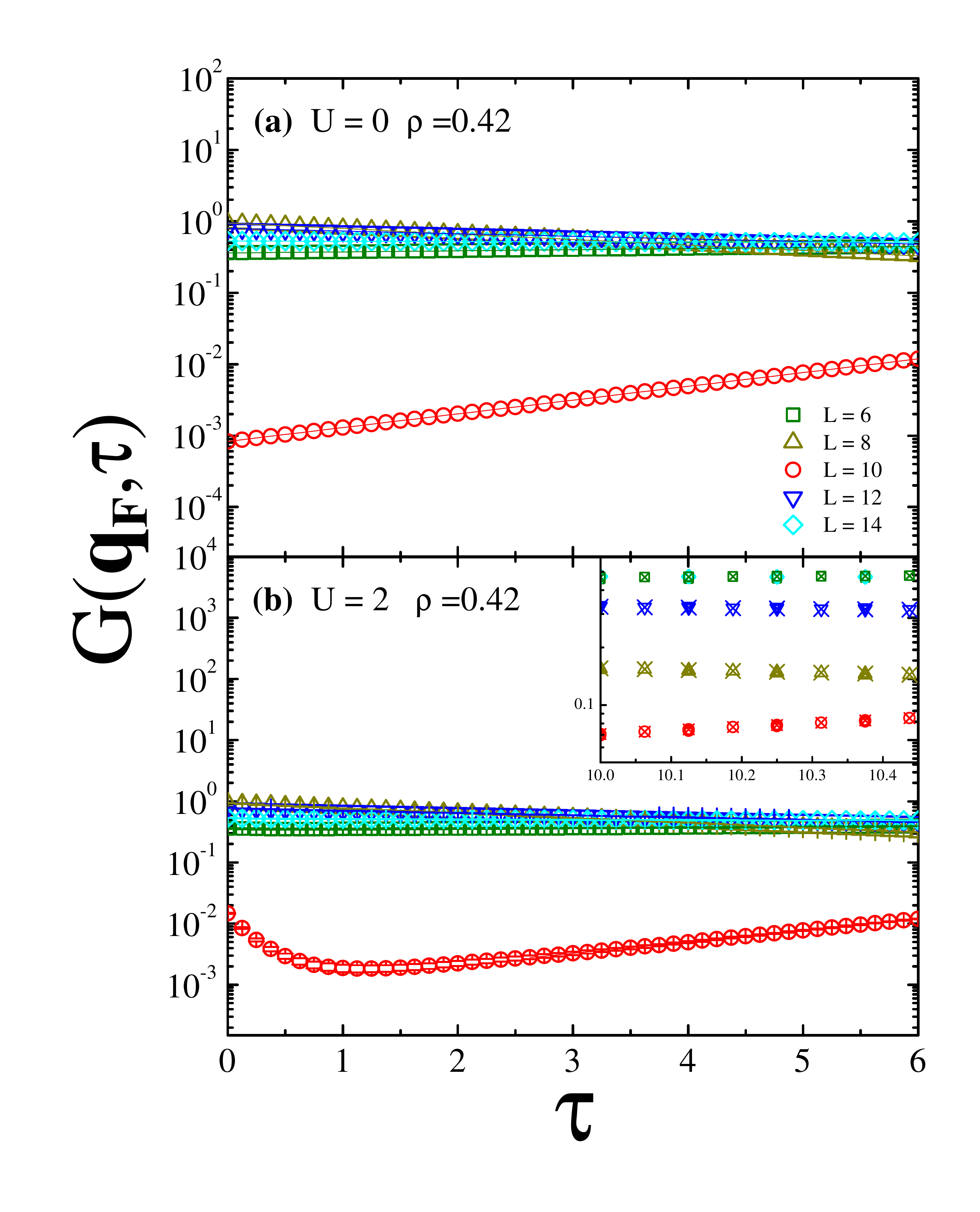}
\vspace{-0.2cm}
\caption{(Color online)
Same as Fig.~\ref{fig:G(q,tau)rho1}, but now for the electronic density $\rho=0.42$.
\vspace{-0.5cm}
}
\label{fig:G(q,tau)rho042} 
\end{figure}
\begin{figure*}[t]
\vspace{-1.0cm}
{\centering\resizebox*{17.5cm}{!}{\includegraphics*{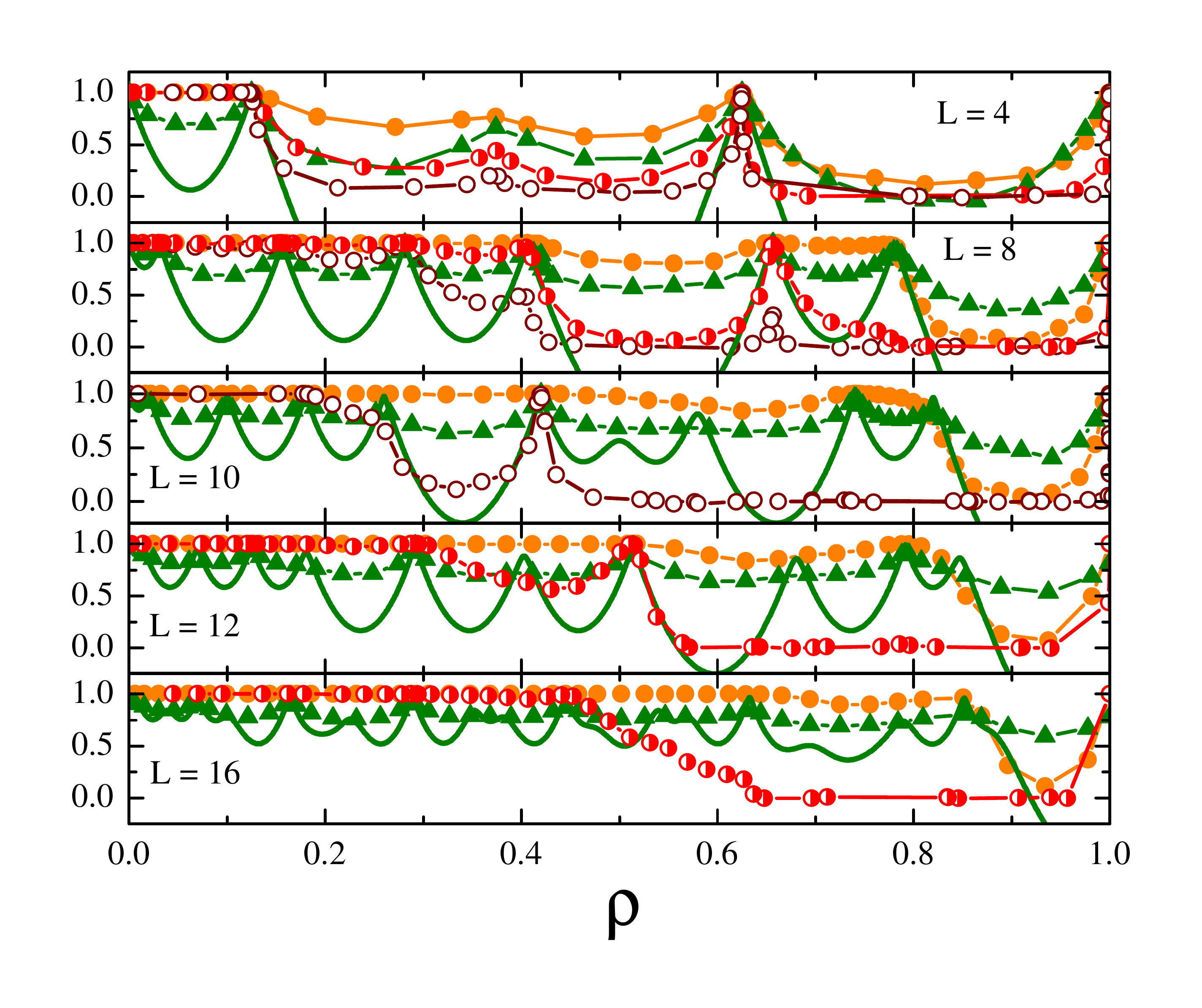}}}
\vspace{-0.7cm}
\caption{(Color online) Average sign (circles) of the fermionic determinant and $\tilde\kappa$ [(green)
thick line for $U=0$, and (green) triangles for $U=2$; see text for the definition of $\tilde\kappa$] as functions of electronic density, for different system sizes. Filled, half-filled, and empty
circles respectively denote $U=2$, 3 and 4; data for $U=0$ are with $\beta=30$, while for $U\neq0$ data are with $\beta=16$.
For the sake of clarity, error bars were omitted, since they are smaller than data points.
}
\label{fig:sign_kappa} 
\end{figure*}

\section{Single-particle excitation gap}
\label{sec:gap}

Another quantity used to infer the transport properties of the system is the single-particle excitation gap $\Delta_{sp}(\textbf{q})$, which is the minimum energy necessary to extract one fermion from the system, and is, essentially, related to the gap measurable in photoemission experiments.
It can be obtained from the imaginary-time--dependent Green's function in reciprocal space, which for large $\tau$ decays exponentially, i.e., $G(\mathbf{q}_F,\tau)\sim e^{-\Delta_{sp}(\mathbf{q}_F)\tau}$ (see, e.g., Ref.\ \onlinecite{Meng10}).
We can therefore obtain $\Delta_{sp}$ through fits of QMC data for the Green's function, calculated at the Fermi wavevector for the electronic densities of interest.
Figure \ref{fig:G(q,tau)rho1} shows the imaginary-time dependence of the Green's function for the half-filled case.
In the upper panel, the absence of a decay in the non-interacting case is a signature of a metallic state, while the exponential decay in the lower panel results from a finite gap.
The inset in (b) compares data obtained for two values of $\Delta\tau$: the time-dependent Green's functions lie on the same exponential curve, which illustrates that this quantity is also negligibly dependent on the  $\Delta\tau$ used.
The size dependence of the gap is shown in Fig.\ \ref{fig:gapsprho1}, for different values of $U$;
for $U=2$, one also sees that data for a smaller $\Delta\tau$ lie on the same curve.
The limiting (i.e., $L\to\infty$) value of $\Delta_{sp}$ increases from zero with increasing $U$, as expected; it is again clear that the value of $\Delta\tau$ does not influence this extrapolation procedure.

Figure \ref{fig:G(q,tau)rho042} shows data for the Green's function for the density $\rho=0.42$.
In the non-interacting case, and discarding the data for $L=10$, we see that the slope decreases as $L$ increases, leading to a vanishing gap as $L\to \infty$, as one would expect for a metallic system; the data for $L=10$ are completely off the mark, again as a result from the closed-shell density for this $L$.
For the interacting case [Fig.\ \ref{fig:G(q,tau)rho042}(b)], the Green's function for $L\neq 10$ behaves in a way similar to that for the free case; again, the $L=10$ case behaves completely differently from the others, bearing a negative gap as the signature of the closed-shell problem.
In this respect, it is interesting to have in mind that the single-particle excitation gap provides a very clear indication that a closed-shell incident is at play for a given combination of $\rho$ and $L$.
For completeness, we note that, similarly to half filling (Fig.\ \ref{fig:G(q,tau)rho1}), the dependence with $\Delta\tau$ is negligible.

\section{The Minus-sign Problem}
\label{sec:sign}

In the present formulation of the QMC method, once the fermionic degrees of freedom are traced out, the role of Boltzmann factor in the partition function is played by the product of two determinants; see, e.g., Refs.\ \onlinecite{Hirsch85,Loh90,dosSantos03}.
Since one cannot guarantee that this product is positive definite for each configuration of the auxiliary fields, the averages are carried out in the ensemble of positive Boltzmann weights, at the expense of having to divide these averages by the average sign of the product of determinants, $\ave{sign}$.
Therefore, when $\ave{sign}$ becomes significantly smaller than 1, the average values of most quantities of interest become meaningless: this is the infamous `minus-sign problem'.
It should be noted that other implementations of the QMC method also run into similar problems; see, e.g., Ref.\ \onlinecite{vdl92}.

This problem has eluded a variety of attempts of solution proposed over the years; see, e.g., Ref.\ \onlinecite{dosSantos03} for a partial list of references.
For instance, once realized that simply ignoring the negative sign leads to serious discrepancies,\cite{Loh90} attempts to use different Hubbard-Stratonovich transformations turned out to be fruitless;\cite{Batrouni90,Batrouni93}
the minus-sign problem has been alleviated with implementations of QMC constraining the sampling process,\cite{Zhang95,Zhang97,Cosentini98,Chang08} from which a ground-state wave function is obtained.
Other frameworks have been proposed to improve the sign problem,\cite{Mak98,Dikovsky01,Umrigar07} but systematic implementations comparing results for, e.g., correlation functions in the Hubbard model are, as far as we know, still unavailable.

More recently, arguments have been given\cite{Troyer05} suggesting that there is no generic solution to the sign problem; instead, in the most favorable scenario, one may find special solutions for specific models.\cite{Troyer05}

In view of this, it is imperative to gather as much information as possible about $\ave{sign}$.
With this in mind, we define a quantity, $\tilde\kappa\equiv 1-\rho^2 \kappa$, directly related to the compressibility $\kappa$ defined in Sec.\ \ref{sec:kappa}.
Figure \ref{fig:sign_kappa} shows that $\tilde\kappa$ reaches the value 1 at the densities corresponding to `closed-shell', as already discussed.
In the same figure we also show $\ave{sign}$ as a function of $\rho$: interestingly, we see that it tracks $\tilde\kappa$, in the sense that, at least for $U\leq2$, it is harmless at densities such that $\tilde\kappa\approx 1$ ($\kappa\approx 0$), but it can be seriously deleterious to the QMC averaging process when the system is more compressible, especially at larger values of $U$.
Since a larger compressibility, in turn, corresponds to stronger density fluctuations, one may conclude that these are inherently linked with the minus-sign problem.
It is worth noticing that improvements on convergence have been achieved within both projector\cite{Furukawa91,Furukawa92} and fixed-node\cite{Cosentini98} QMC simulations if closed-shell configurations are used as initial states; in addition, in Ref.\ \onlinecite{Furukawa91} it was also pointed out that the choice of closed-shell initial states led to larger $\ave{sign}$ than when open-shell initial states were taken.
On the other hand, shell effects have also disrupted the density dependence needed in the search for phase separation in the $t$-$J$ model.\cite{Hellberg97,Hellberg00}
Thus, while indications of an interplay between closed-shell and the minus-sign problem have been suggested in the past, Fig.\ \ref{fig:sign_kappa} presents the first systematic evidence of this connection.

\begin{figure}[t]
\centering
\includegraphics[width=9.0cm,angle=0]{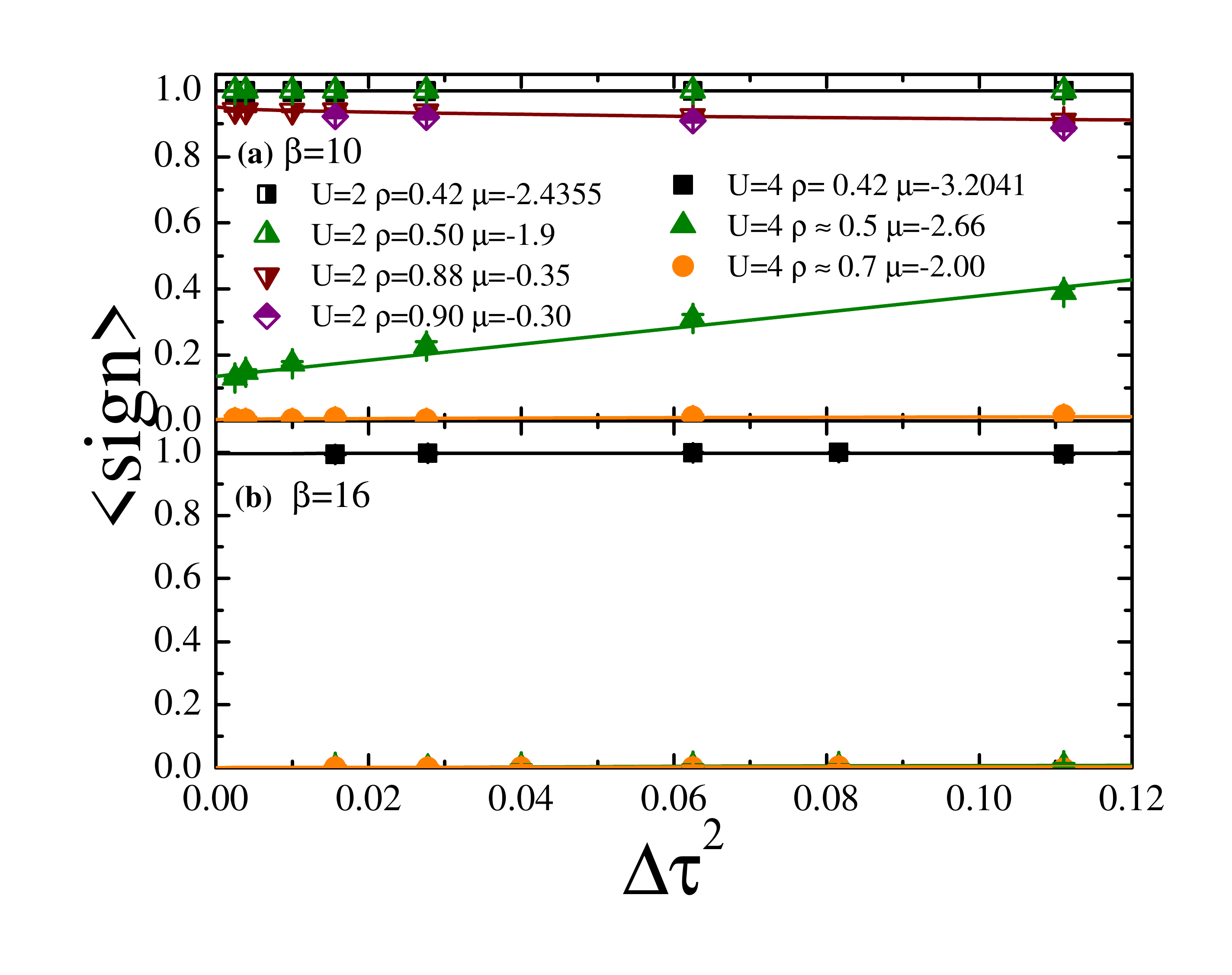}
\caption{
(Color online)
Average sign of the fermionic determinant as a function of the square of Suzuki-Trotter `time' interval, for a $10\times 10$ lattice, with (a) $\beta=10$, and (b) $\beta=16$.
Black squares and (green) up triangles respectively correspond to the closed-shell density $\rho=0.42$, and  to $\rho\approx0.5$; half-filled and filled symbols respectively correspond to $U=2$ and 4.
}
\label{fig:sign_dtau} 
\end{figure}

\begin{figure}[h!]
\centering
\includegraphics[width=8.5cm,angle=0]{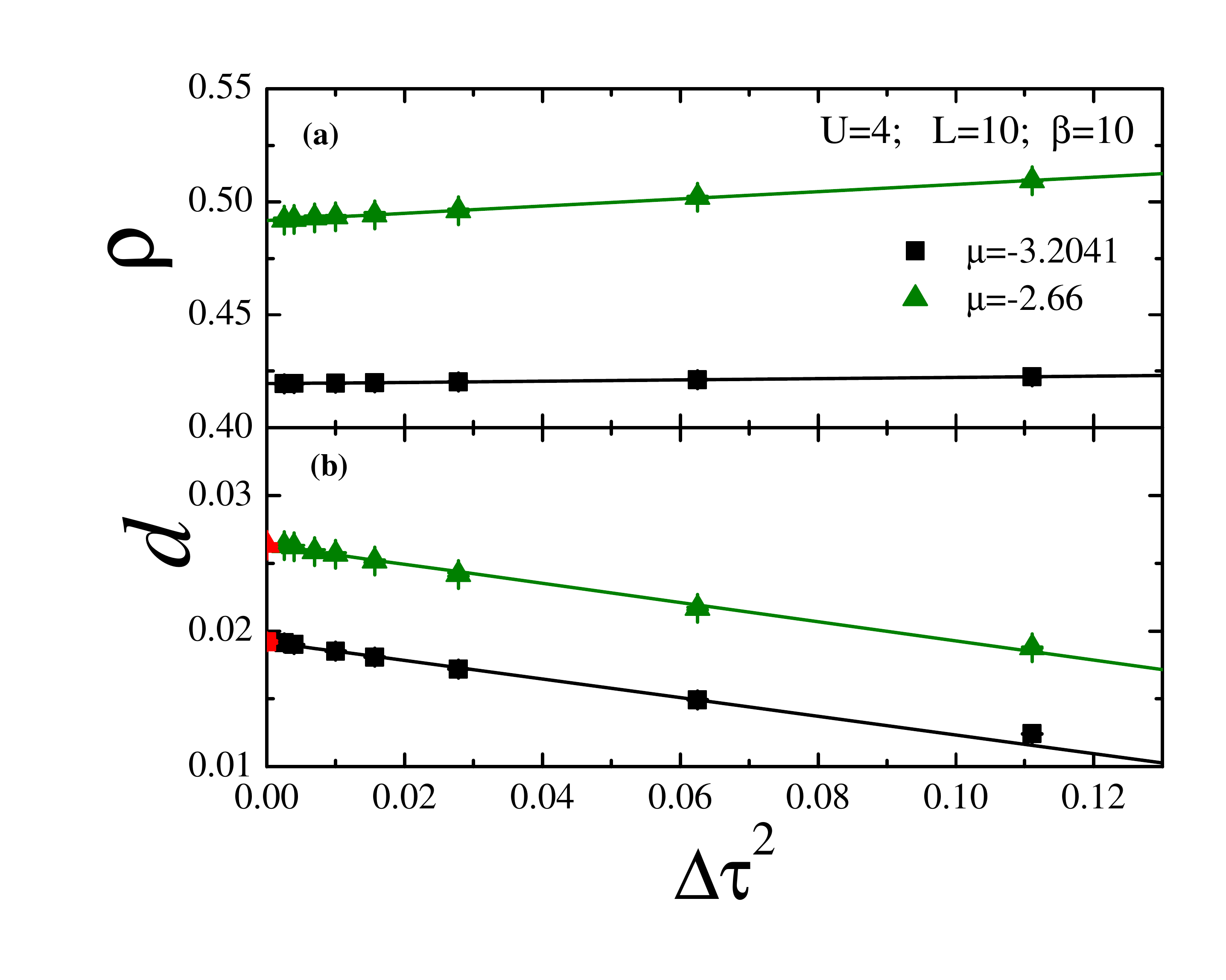}
\caption{
(Color online)
Dependence of average values of (a) electronic density, and (b) double occupancy (see text) with the square of the Suzuki-Trotter `time' interval, for two values of the chemical potential; $U$, $\beta$, and lattice size are fixed. The extrapolated values, obtained from the fitting of a straight line through all points, are shown in red at $\Delta\tau=0$.}
\label{fig:rho-docc-dtau2} 
\end{figure}

It is also instructive to examine the behavior of $\ave{sign}$ with $\Delta\tau$.
Figure \ref{fig:sign_dtau} compares data for one lattice size, $L=10$, but for different values of $U$,
$\beta$, and the chemical potential $\mu$.
For $\beta=10$, we see that for the closed shell configuration, $\rho=0.42$, $\ave{sign}\approx 1$ for all $\Delta\tau$ in the range considered, for both $U=2$ and 4; this feature is maintained when $\beta$ is increased to 16, illustrating the harmlessness of $\ave{sign}$ at the closed-shell density.
Doping slightly away, e.g., for an open shell configuration with $\rho\approx 0.5$, $\ave{sign}$ remains almost independent of $\Delta\tau$ for $U=2$, but acquires a significant dependence for $U=4$, leading to low values for small $\Delta\tau$; for $\beta=16$, $\ave{sign}\approx 0$ for all $\Delta\tau$ in the relevant range.
Worse still, for $U=4$ and $\rho\approx 0.7$, $\ave{sign}$ is very close to zero for all values of $\Delta\tau$ considered, for both $\beta=10$ and 16; this should not come as a surprise, since Fig.\ \ref{fig:sign_kappa} shows that $\tilde\kappa$ vanishes near this density for the $10\times 10$ lattice.

In Figure \ref{fig:rho-docc-dtau2} we display the dependence of two average local quantities with $\Delta\tau^2$, for two fixed values of the chemical potential, and for $U=4$ and $\beta=10$.
Notwithstanding the fact that systematic errors of order $\Delta\tau^2$ are expected as a result of the Suzuki-Trotter decomposition, panel (a) shows that for $\mu\approx -3.2$, the proportionality constant is quite small, so that $\rho=0.42$ over the whole range of $\Delta\tau$; in the open shell case, for which $\ave{sign}$ deteriorates with decreasing $\Delta\tau$ (see Fig.\ \ref{fig:sign_dtau}), the dependence of $\rho$ with $\Delta\tau^2$ is noticeable.
By contrast, the lower panel shows that the double occupancy,
\begin{equation}
d\equiv \ave{n_{\iv\up}n_{\iv\dn}},
\label{eq:docc}
\end{equation}
follows the expected linear dependence with $\Delta\tau^2$ in both cases.
This indicates that whenever $\ave{sign}$ is strongly dependent on $\Delta\tau$, one can still obtain meaningful averages by using solely the data for the largest values of $\Delta\tau$ to extrapolate towards $\Delta\tau=0$; though with less confidence, the same procedure could be adopted for $\beta=16$.


\section{Conclusion}
\label{sec:concl}

In conclusion, we have thoroughly examined the behavior of several quantities, obtained through QMC simulations at finite temperatures for the homogeneous Hubbard model on the square lattice, and commonly used to locate insulating behavior.
Our results show that `closed-shell' effects, which introduce important (though artificial) gaps in the spectrum, may lead to {\em false} insulating behavior of the compressibility, of the conductivity, and of the charge gap at certain combinations of occupation and linear lattice size, $L$; in situations in which a long series of lattice sizes cannot be obtained, this may jeopardize extrapolations towards $L\to\infty$.
We have also assessed corrections to the dc-conductivity, which are neglected when a Laplace transform  is avoided through a simplifying prescription, and found that the latter is not generically valid due to the absence of a sufficiently small energy scale in the problem; though quite appealing, fittings to experimental data with the conductivity thus obtained should be avoided.
The Drude weight, on the other hand, suffers from more controllable finite-size and finite-temperature effects.
At half filling, and at a fixed low temperature, it vanishes with a power law in $1/L$, the exponent of which depends on $U$; away from half filling, the Drude weight is only weakly dependent on either temperature and system size, being free from the spurious behavior found in other quantities.
Therefore, amongst all quantities discussed here, the Drude weight is certainly the most reliable one to use in situations for which the data are limited to a restricted set of system sizes.

In addition, we have also presented numerical evidence showing that the sign of the fermionic determinant tracks the compressibility: for densities at which the system is `incompressible', as a result of a gap due to the finiteness of the lattice,  $\ave{sign}\approx 1$, at least for $U\leq 2$.
However, in-between two successive `incompressible' densities, $\ave{sign}$ deteriorates steadily as $U$ increases.
This behavior is suggestive that strong density fluctuations may be linked to the `minus-sign problem'.
We have also investigated the influence of the imaginary-time interval $\Delta\tau$ on the behavior of $\ave{sign}$ and of some (local) average quantities.

All analysed quantities  can be fitted to a linear dependence with $\Delta\tau^2$, as expected from the Suzuki-Trotter discretization, although at the closed-shell density, the slopes for both $\ave{sign}$ and the density $\rho$ are very small.
The $\Delta\tau^2$dependence  is indicative that for some densities, one can confidently use data for `large' $\Delta\tau$ (i.e., those leading to $\ave{sign}\gtrsim 0.5$) to perform extrapolations (towards $\Delta\tau\to0$) of average values.

\begin{acknowledgments}

We are grateful to R.T. Scalettar for useful discussions. Financial support from the Brazilian Agencies CNPq, CAPES, and FAPERJ is  gratefully acknowledged.

\end{acknowledgments}


\begin{thebibliography}{99}
\bibitem{alpha}
As a technical note, we should mention that the parameter $\alpha$ in Ref.\ \onlinecite{Sandvik98} is adjusted dynamically in such a way that the argument of the exponential in the likelihood function vanishes.
\end{thebibliography}
\end{document}